\documentclass[pra,twocolumn,aps,superscriptaddress,showpacs,epsfig]{revtex4}
\usepackage{epsfig}
\usepackage{graphicx}
\usepackage{amsmath}

\newcommand{\beq}{\begin{eqnarray}}
\newcommand{\eeq}{\end{eqnarray}}
\def\be{\begin{equation}}
\def\ee{\end{equation}}
\def\ba{\begin{eqnarray}}
\def\ea{\end{eqnarray}}

\begin{document}
\title{Two-mode Bose-Einstein condensate in a high-frequency driving
field that directly couples the two modes}
\author{Qi Zhang}
\affiliation{Department of Physics and Center for Computational
Science and Engineering, National University of Singapore,117542,
Republic of Singapore}
\author{Peter H\"anggi}
\affiliation{Department of Physics and Center for Computational
Science and Engineering, National University of Singapore,117542,
Republic of Singapore} \affiliation{Institut f\"ur Physik,
Universit\"at Augsburg, Universit\"atsstra{\ss}e 1, D-86135
Augsburg, Germany}
\author{Jiangbin Gong}
\email{phygj@nus.edu.sg} \affiliation{Department of Physics and
Center for Computational Science and Engineering, National
University of Singapore,117542, Republic of Singapore}
\affiliation{NUS Graduate School for Integrative Sciences and
Engineering, Singapore
 117597, Republic of Singapore}
\date{\today}
\begin{abstract}
A two-mode Bose-Einstein condensate coupled by a high-frequency
modulation field is found to display rich features. An effective
stationary Hamiltonian approach reveals the emergence of additional
degenerate eigenstates as well as new topological structures of the
spectrum. Possible applications, such as the suppression of
nonlinear Landau-Zener tunneling, are discussed. An interesting
phenomenon, which we call ``deterministic symmetry-breaking
trapping" associated with separatrix crossing, is also found in an
adiabatic process. \end{abstract}


\maketitle

\section{introduction}
Significant research efforts have been devoted to the nonlinear
dynamics of interacting cold atoms, e.g., a Bose-Einstein condensate
(BEC) in an external driving field. One main motivation is to
understand how we can actively control the nonlinear dynamics and
how the self-interaction of cold atoms can be used to simulate
fundamental models \cite{lattice}. Examples include the control of
the BEC self-trapping \cite{exp-trapping,holthaus1,luo}, effective
turning-off of the self-interaction \cite{Zoller}, controlled
Mott-insulator transitions associated with a BEC in an optical
lattice \cite{phase-transitions}, stabilization of bright BEC
solitons by an oscillating magnetic field tuned close to the
Feshbach resonance \cite{ueda}, as well as production of ultracold
molecules using stimulated Raman adiabatic passage
\cite{photoassociation,itin}. On a deeper level, BEC systems offer a
useful tool to explore more aspects of  many-body systems. In
particular, the dynamics of a BEC in the large-particle-number limit
is described by a mean-field nonlinear Schrodinger equation
(Gross-Pitaeviskii equation). The resulting nonlinearity often
challenges existing theories for linear systems. For example, the
adiabatic following of a two-mode nonlinear system with an external
field may necessarily break down \cite{Wu}.


Here we aim to examine how an adiabatic Landau-Zener (LZ) tunneling
process of a BEC may be manipulated by an external driving field,
thus extending an earlier study for linear systems \cite{Hanggi}. As
a second motivation on a more fundamental level, we shall expose
some  nonlinear dynamics phenomena that do not exist in the
mean-field dynamics of a BEC under field-free conditions.
Specifically, we consider a two-mode BEC under a high-frequency
field that directly couples the two modes (hence called
``off-diagonal" driving below). A biased
static field is also considered for LZ processes. 

It is well-known that a high-frequency ``diagonal modulation" (e.g.,
high-frequency tilting of a double-well potential) can only re-scale
the natural two-mode coupling strength
\cite{phase-transitions,modulation1,modulation2,modulationLiu}.
However, our findings below for a different type of high-frequency
modulation are different.  In particular, we show (i) that a
high-frequency ``off-diagonal" driving field can induce, in effect,
additional nonlinear terms in the mean-field equations of motion,
thus offering a control ``knob" to tune different nonlinear terms
and simulate some systems not considered before; (ii) that different
topological structures of the eigen-spectrum of the system can be
generated and tuned by the driving field, also leading to additional
degenerate eigenstates. We then suggest to use a high-frequency
driving field to realize the complete suppression of nonlinear LZ
tunneling. In analyzing the adiabatic dynamics of a driven nonlinear
system, we also find and qualitatively explain a phenomenon, called
``deterministic symmetry-breaking trapping" associated with
separatrix crossing. 
\section{Two-mode system under off-diagonal modulation}
The nonlinear two-mode system under off-diagonal modulation is
described by
\begin{equation}\label{Hamiltonian}
{H}(t)=\frac{1}{2}\left(\begin{array}{cc}\gamma+c(|b|^{2}-|a|^{2})&\Delta_{0}+A\sin(\omega t)\\
\Delta_{0}+A\sin(\omega
t)&-\gamma-c(|b|^{2}-|a|^{2})\end{array}\right);
\end{equation}
where $\gamma$ denotes an external energy bias,  $|a|^{2}$ and
$|b|^{2}$ represent occupation probabilities for the two modes, $c$
characterizes the nonlinear atom-atom interaction, and $\Delta_{0}$
denotes the static coupling between the two modes.
We put $\hbar=1$
throughout this work.
There are a number of possibilities to experimentally realize this
Hamiltonian. For example, one may consider a BEC in a double-well
potential, with the height of the potential barrier periodically
modulated, or a BEC in an optical lattice occupying two bands, with
the well-depth of the optical lattice periodically modulated. In
principle, these procedures should be achievable, considering
previous experiments on two-mode BEC's \cite{double-well,two-band}.
What might be even more feasible in realizing this two-mode system
under off-diagonal modulation is to consider the internal states of
a BEC, such as $^{87}Rb$ \cite{Rb}, where there exist two internal
states separated by a relatively large hyperfine energy. Then, the
energy bias $\gamma$ can be effectively realized by the detuning of
the coupling field from the resonance and the off-diagonal
modulation may be achieved by modulating the intensity of the
coupling field. Considering recent studies of two-mode nonlinear
Schr\"{o}dinger equations using nonlinear optical waveguides (for
example, see in Ref. \cite{luo}), it might be also possible to
realize our system in nonlinear optics.

Consider first the non-driven case, i.e. $A=0$. Then ${H}(t)$
reduces to the standard model of nonlinear LZ tunneling \cite{Wu}.
Therein the eigen-spectrum diagram as a function of $\gamma$ is
known to display a loop structure at the tip of the lower (upper)
level for $c>\Delta_{0}$ ($c<-\Delta_{0}$). Such a loop structure,
absent in linear systems, directly leads to a nonzero LZ transition
probability even when $\gamma$ changes adiabatically. As shown
below, new system properties emerge if the driving field is turned
on. Without loss of generality we will restrict ourselves to the
$c>0$ case, which requires an attractive interaction for bosons in a
double-well potential or a repulsive interaction for bosons in two
energy bands of an optical lattice.

In the general case of $A\neq 0$ with $\omega\gg
\gamma,c,\Delta_{0}$, it is found that $|a|^{2}$ ($=1-|b|^{2}$) also
oscillates at the frequency $\omega$. To expose possibly new physics
hidden in the oscillations, another pair of wave function parameters
$(a',b')$ are found to be very useful, i.e.,
\begin{eqnarray} \label{relation} \nonumber
a'=\frac{a+b}{2}e^{-i\frac{A}{2\omega}\cos(\omega
t)}+\frac{a-b}{2}e^{i\frac{A}{2\omega}\cos(\omega t)},\\
b'=\frac{a+b}{2}e^{-i\frac{A}{2\omega}\cos(\omega
t)}-\frac{a-b}{2}e^{i\frac{A}{2\omega}\cos(\omega t)}.
\end{eqnarray}
Their equations of motion are given by
\begin{eqnarray}\label{NewHamiltonian}
\nonumber i\frac{da'}{dt}&=&
\frac{1}{2}[\gamma\cdot \cos(\theta)+c\cdot \cos^{2}(\theta)(|b'|^{2}-|a'|^{2})\\
\nonumber&&-ic\cdot \sin(\theta)\cos(\theta)(a'^{*}b'-a'b'^{*})]a'\\
\nonumber&&+\frac{1}{2}[\Delta_{0}+i\gamma \cdot \sin(\theta)+c\cdot \sin^{2}(\theta)(a'^{*}b'\\
\nonumber&&-a'b'^{*})-ic\cdot
\sin(\theta)\cos(\theta)(|a'|^{2}-|b'|^{2})]b'\nonumber \\ \nonumber
\\
 i\frac{db'}{dt}&=&\frac{1}{2}[\Delta_{0}-i\gamma \cdot
\sin(\theta)-c\cdot \sin^{2}(\theta)(a'^{*}b'\nonumber \\
\nonumber&&-a'b'^{*})+ic\cdot \sin(\theta)\cos(\theta)(|a'|^{2}-|b'|^{2})]a'\\
\nonumber&&+\frac{1}{2}[-\gamma \cdot
\cos(\theta)-c\cdot \cos^{2}(\theta)(|b'|^{2}-|a'|^{2})\\
&&+ic\cdot \sin(\theta)\cos(\theta)(a'^{*}b'-a'b'^{*})]b',
\end{eqnarray}
where $\theta\equiv \frac{A}{\omega}\cos(\omega t)$.
Along with previous studies that focused on high-frequency driving
fields
\cite{phase-transitions,modulation1,modulation2,modulationLiu}, we
consider now sufficiently large $\omega$, such that the oscillation
in $\theta$ is much faster than the natural time scale of the system
as characterized by $\Delta_{0}$, $\gamma$, and $c$ (numerically, we
find that the regime of $\omega>10\Delta_{0}$, $\omega>10c$, and
$\omega>10\gamma_0$, where $\gamma_0$ is the initial value of
$|\gamma|$ that is sufficiently large to ensure the LZ dynamics, can
be safely regarded as a high-frequency regime; experimentally, a
high-frequency driving field should not interfere with the two-mode
descriptions). Then Eq. (\ref{NewHamiltonian}) can be significantly
reduced by considering the averages of $(a',b')$ over $2\pi/\omega$.
Speaking more rigorously, upon the large frequency condition, a
zeroth-order approximation of a ``$1/\omega$" expansion can be used
to yield
\begin{eqnarray}\label{Newmotion}
\nonumber i\frac{da'}{dt}&=&\frac{1}{2}[\gamma' +c_{Z}
(|b'|^{2}-|a'|^{2})]a'\\
\nonumber&&+\frac{1}{2}[\Delta_{0}+c_{Y}
(a'^{*}b'-a'b'^{*})]b'\\
\nonumber i\frac{db'}{dt}&=&\frac{1}{2}[\Delta_{0}-c_{Y}(a'^{*}b'-a'b'^{*})]a'\\
&&+\frac{1}{2}[-\gamma'-c_{Z}(|b'|^{2}-|a'|^{2})]b',
\end{eqnarray}
where $\gamma'=\gamma \langle\cos(\theta)\rangle_{T}=\gamma
J_{0}(A/\omega)$ \cite{modulation2}, $c_{Z}=c
\langle\cos^{2}(\theta)\rangle_{T}=c[\frac{1+J_{0}(2A/\omega)}{2}]$,
and
$c_{Y}=c\langle\sin^{2}(\theta)\rangle_{T}=c[\frac{1-J_{0}(2A/\omega)}{2}]$
($J_{0}$ is the zeroth order Bessel function of the first kind).
Evidently, these newly defined parameters reflect the action of the
high-frequency driving field. We stress that the validity of this
kind of high-frequency approximation has been checked numerically
and has been used in many situations.

Equation (\ref{Newmotion}) no longer explicitly contains a
time-dependent field.  We can then define an effective static
Hamiltonian $H_{\text{eff}}$ that generates Eq. (\ref{Newmotion}).
That is,
\begin{equation} \label{effective}
{H}_{\text{eff}}=\frac{1}{2}\left(\begin{array}{cc}\gamma+c_{Z}(|b|^{2}-|a|^{2})&\Delta_{0}+c_{Y}(a^{*}b-ab^{*})\\
\Delta_{0}-c_{Y}(a^{*}b-ab^{*})&-\gamma
-c_{Z}(|b|^{2}-|a|^{2})\end{array}\right),
\end{equation}
where, for simplicity, we have replaced $a'$ by $a$, $b'$ by $b$,
and so on. If we compare $H_{\text{eff}}$ with the original
Hamiltonian in Eq. (\ref{Hamiltonian}) for $A=0$, one sees that the
nonlinear parameter $c_{Z}$ can be regarded as a rescaled parameter
$c$, and the nonlinear term containing $c_{Y}$ is new. In addition,
the ratio of $c_{Z}$ and $c_{Y}$ is given by
$[1+J_{0}(2A/\omega)]/[1-J_{0}(2A/\omega)]$,  easily adjustable by
choosing different $\omega$ and $A$.

The effective Hamiltonian in Eq.(\ref{effective}) can be recognized
as the one describing a single spin in a biaxial crystal field, with
the $c_{Y}$ ($c_{Z}$) term describing the anisotropy in the $Y$
($Z$) direction. Indeed, ${H}_{\text{eff}}$ can also be written as
$H_{\text{spin}}=\gamma
S_{Z}+\Delta_{0}S_{X}-c_{Z}S_{Z}^{2}-c_{Y}S_{Y}^{2}$, where
$S_{Z}=\frac{|a|^{2}-|b|^{2}}{2}$, $S_{X}=\frac{a^{*}b+b^{*}a}{2}$
and $S_{Y}=\frac{a^{*}b-b^{*}a}{2i}$. The corresponding
second-quantization Hamiltonian exactly describing the quantum
system with $N$ bosons on the two modes is given by
\begin{eqnarray}\label{quantized}
\nonumber
\hat{H}_{Q}=&\gamma\frac{(\hat{a}^{\dag}\hat{a}-\hat{b}^{\dag}\hat{b})}{2}+\Delta_{0}\frac{(\hat{a}^{\dag}\hat{b}+\hat{a}\hat{b}^{\dag})}{2}
-\frac{c_{Z}}{N}(\frac{\hat{a}^{\dag}\hat{a}-\hat{b}^{\dag}\hat{b}}{2})^{2}\\
&+\frac{c_{Y}}{N}(\frac{\hat{a}^{\dag}\hat{b}-\hat{b}^{\dag}\hat{a}}{2})^{2}.
\end{eqnarray}

\begin{figure}[t]
\begin{center}
\vspace*{-0.5cm}
\par
\resizebox *{9cm}{9cm}{\includegraphics*{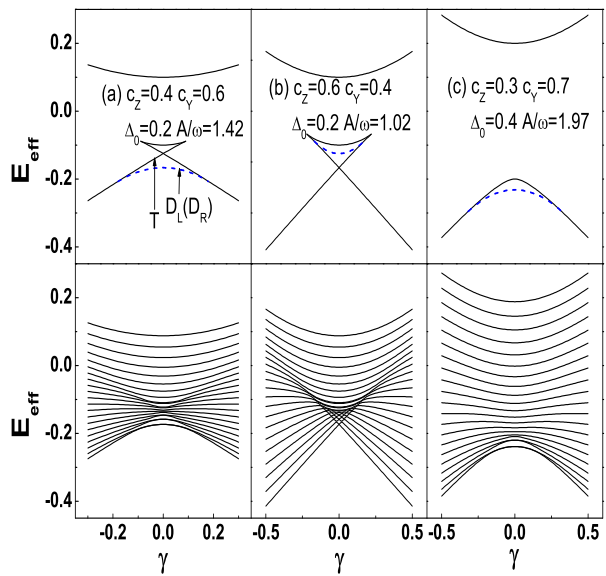}}
\end{center}
\par
\vspace*{-0.5cm} \caption{Upper panels: Level structures of the
stationary effective Hamiltonian $H_{\text{eff}}$ [see Eq.
(\ref{effective})], as a function of $\gamma$. The dashed lines
denote the new mean-field levels that are absent in a non-driven
two-mode BEC. Symbols $T$, $D_{R}$, and $D_{L}$ indicate how the
involved levels are connected with the phase space structures shown
in Fig. 2. Bottom panels: Parallel results in a fully quantum
treatments for $N=20$.} \label{fig1}
\end{figure}

\section{Detailed Results}
We now present in Fig. 1 the eigen-spectrum of $H_{\text{eff}}$ as a
function of $\gamma$.  Evidently, the typical level structures (such
as the loop structure) for a nonlinear LZ tunneling model \cite{Wu}
are also possessed by our system.  On top of that,  additional
mean-field eigenstates (dashed lines) that are absent in a
non-driven case also emerge through level bifurcations. The new
eigenstates are directly caused by the $c_{Y}$ term induced by the
driving field. In particular, if a loop structure exists and if
$c_{Y}<c_{Z}$, then the additional level lies inside the loop, as
shown in Fig. 1(b); and if $c_{Y}>c_{Z}$ and $c_{Y}>\Delta_{0}$,
then level bifurcation takes place on the lowest branch and the
additional level can be below the loop structure, as shown in Fig.
1(a). Figure 1(c) shows that the additional level may also exist in
the absence of a loop structure. In the bottom panels of Fig. 1, we
also show fully quantum mechanical levels calculated from Eq.
(\ref{quantized}) for $N=20$. The results confirm that the
additional eigenstates we obtain on the mean-field level do have
physical implications for fully quantum levels, even in the cases
with a not very large $N$.

\begin{figure}[t]
\begin{center}
\vspace*{-0.5cm}
\par
\resizebox *{10.6cm}{8.6cm}{\includegraphics*{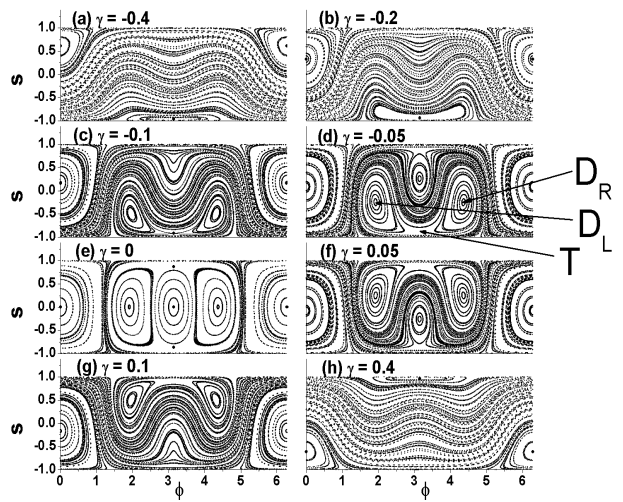}}
\end{center}
\par
\vspace*{-0.5cm} \caption{Phase space structures of $H_{c}$ defined
in Eq. (7), for $\Delta_{0}=0.2$; and $c_{Z}=0.4$, $c_{Y}=0.6$ (or
equivalently, $c=1$, $A/\omega=1.42$). ``T" denotes an unstable
fixed point with $\phi=\pi$. $D_{R}$ and $D_{L}$ denote two stable
fixed points on the right and on the left.} \label{fig2}
\end{figure}

Let us now examine Eq. (\ref{Newmotion}) from a phase space
perspective, by mapping the mean-field trajectories of Eq.
(\ref{Newmotion}) to that of a well-defined classical Hamiltonian
system. The associated phase space can be defined in terms of $s$
and $\phi$, where $\phi = \phi_{b}-\phi_{a}$
\cite{Liu,modulationLiu}, $s=|b|^{2}-|a|^{2}$, with
$a=|a|e^{i\phi_{a}}$ and $b=|b|e^{i\phi_{b}}$. Using this pair of
canonical variables, the involved classical Hamiltonian is:
\begin{eqnarray}\label{classical}
H_{c}&=&\frac{1}{2}[-\gamma
s-\frac{c_{Z}}{2}s^{2}+\Delta_{0}\sqrt{1-s^{2}}\cos(\phi) \nonumber
\\ &-&\frac{c_{Y}}{2}(1-s^{2})\sin^{2}(\phi)].
\end{eqnarray}
The nonlinear eigenstates of $H_{\text{eff}}$ now become fixed
points in the phase space of $H_{c}$. Figure 2 displays phase space
portraits of $H_{c}$ for the parameters used in Fig. 1(a), for
several values of $\gamma$ covering the regime of level bifurcation.
In particular, the lower parts of Fig. 2(b), 2(c) and 2(d) near
$\phi=\pi$ clearly show the splitting of one fixed point into three
fixed points, thus associating the level bifurcation in Fig. 1 with
the splitting of a fixed point. Because both the elliptic (stable)
fixed points (marked by ``$D_{R}$" and ``$D_{L}$" in Fig. 2) yield
the dashed line in Fig. 1(a), the additional level shown in Fig. 1
in fact denotes two-fold degenerate eigenstates. By contrast, the
hyperbolic (unstable) fixed point marked by ``T" in Fig. 2(d) yields
the level right above the degenerate eigenstates [also marked by
``T" in Fig. 1(a)]. In Fig. 2(f) and 2(g), the above-mentioned three
fixed points start to merge back to one fixed point, in parallel
with the level merging seen in Fig. 1(a) as $\gamma$ increases
further. Examining the phase space globally, it is also clear that
the number of the fixed points and hence the number of the nonlinear
eigenstates of $H_{\text{eff}}$ can vary from two to six, a clear
sign that the nonlinear dynamics of a driven BEC can be very rich.

It should also be noted that the above-mentioned two-fold degeneracy
occurs in a high-dimensional parameter space. In particular, for
fixed nonlinear parameter $c$ and fixed field parameters $A$ and
$\omega$, the degeneracy can still occur in a two-parameter space of
$\Delta_{0}$ and $\gamma$. This is in contrast to the well-studied
non-driven model of a two-mode BEC where degeneracy occurs only
along a line for fixed $c$, $A$ and $\omega$.

So, how does the additional eigenstate shown in Fig. 1(a) [cases in
Fig. 1(b) and Fig. 1(c) are physically less appealing] affect the
adiabatic dynamics? To answer this question we numerically solve Eq.
(\ref{Newmotion}) for the parameters used in Fig. 1(a), with
$\gamma<<0$ and the initial state put on the lowest level. As
$\gamma$ increases very slowly, the system's state is found to
follow the non-degenerate lowest level up to the bifurcation point.
When $\gamma$ increases beyond the ``phase transition" point where
the new two-fold degenerate level emerges, the two-fold degenerate
level becomes the lowest and the system is found to move along the
new level. As $\gamma$ increases further, the two-fold degenerate
level finally disappears and the system reaches the non-degenerate
lowest level again, thus completing the LZ process. During the
entire process, the system remains at the lowest level available and
no transitions to any upper levels are found. Hence, the new
two-fold degenerate level induced by the driving field offers a
means to circumvent the loop structure and hence totally suppress
the nonlinear LZ transition that is doomed to happen if the two-fold
degenerate state were not there.  To connect this observation of
totally suppressed LZ tunneling in the $(a',b')$ representation [see
Eq. (2)] with the direct observable $(a,b)$ in experiments, note
that (i) initially if $A=0$ then $a=a'$ and $b=b'$, and (ii) if $A$
is switched on slowly enough as compared with $\omega$ but fast
enough as compared with the change rate of $(a',b')$ (characterized
by $\gamma$, $\Delta_{0}$ and $c$), then the initial values of
$(a,b)$ are passed to $(a',b')$.

\begin{figure}[t]
\begin{center}
\vspace*{-1.0cm}
\par
\resizebox *{8.0cm}{6cm}{\includegraphics*{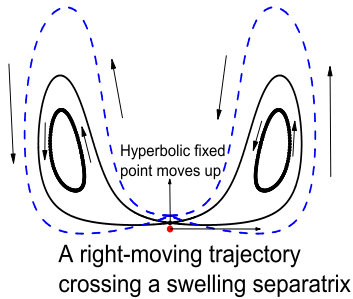}}
\end{center}
\par
\vspace*{-0.5cm} \caption{A classical trajectory falling below an
up-moving unstable fixed point will evolve counter-clockwise around
the associated separatrix (solid line). The separatrix for a
slightly larger $\gamma$ is represented by the dashed line. The
shown trajectory is moving from left to right and will cross the
swelling separatrix on the right and get trapped by the stable fixed
point on the right.} \label{fig3}
\end{figure}

One more aspect of the above nonlinear LZ process remains to be
examined. Because the dashed line in Fig. 1(a) denotes a two-fold
degenerate eigenstate, we should study which state the system will
reside in when it slowly pass the level bifurcation point with an
increasing $\gamma$.  Since the phase space structure shown in Fig.
2 always possesses a mirror symmetry with respect to $\phi=\pi$, one
may intuitively expect that during the LZ process the system is
trapped by either of the two stable points $D_{R}$ or $D_{L}$ in a
random fashion, with equal probability. However, we find that this
picture is incorrect here. Instead, the system is found to be
deterministically trapped by $D_{R}$ [see Fig. 2(d)]. Physically,
this deterministic trapping means that the relative phase between
$a'$ and $b'$ is not random during the LZ process, i.e., it is
robust to small fluctuations in the initial state.  A careful
analysis enables us to explain this intriguing observation
qualitatively. As $\gamma$ increases, the fixed point with
$\phi=\pi$ [see the lower part of Fig. 2(c) and Fig. 2(d)] moves
upwards in the phase space. When this fixed point becomes unstable
[denoted ``T" in Fig. 2(d)], the adiabatic following must break down
and hence the actual trajectory will find itself slightly below the
up-moving fixed point. As such, the trajectory starts to slowly move
counter-clockwise around a separatrix, or from left to right, as
illustrated in Fig. 3.  At the same time, because $\gamma$ is
increasing, the separatrix deforms and swells, and as a result the
trajectory necessarily crosses the separatrix on the right and hence
gets trapped by $D_{R}$. Indeed, if we reverse the adiabatic
process, i.e., passing the bifurcation point with a decreasing
$\gamma$, then one can predict that the system will be trapped by
$D_{L}$, a prediction confirmed numerically.  This counter-intuitive
``deterministic symmetry-breaking trapping" is complementary to a
well-known separatrix-crossing-induced phenomenon, i.e.,
``quasi-random" trapping in classical mechanics (which has been
systematically applied to a few BEC systems \cite{itin}). Encouraged
by the finding here and in efforts to confirm the generality of our
finding, we also studied the adiabatic following dynamics of a
modified rotating pendulum system whose fixed point moves with an
external parameter. Analogous results are also found in this
pendulum case. Hence, it should be of some interest to carry out an
experimental BEC study of the observed symmetry-breaking separatrix
crossing here. Such kind of experiments might also offer new
insights into the validity of the mean-field description of a BEC.

\section{Conclusion}

To conclude, we have theoretically examined the dynamics of a
two-mode BEC driven by a high-frequency driving field that directly
couples the two modes. Based on our results here we expect rich
phenomena in general for a multi-mode BEC under a high-frequency
driving field. Though our results are purely theoretical, it is our
hope that the results here will stimulate future experiments on the
dynamics of a BEC in high-frequency driving fields and on the
control of
nonlinear LZ tunneling dynamics.  

\acknowledgements One of the authors (J.G.) is supported by the
start-up funding (WBS grant No. R-144-050-193-101 and No.
R-144-050-193-133) and the NUS ``YIA'' funding (WBS grant No.
R-144-000-195-123), both from the National University of Singapore.
One of the authors (P.H.) acknowledges support by the DFG, via the
collaborative research grant SFB-486, project A-10.

\end{document}